\input epsf
\documentstyle{mn}
\newif\ifAMStwofonts

\def\tempest%
{\begin{array}{ccc}
1 & 1 & 1 \\
1 & 1 & 1 \\
4 & 3 & 8
\end{array}}
\def\lesssim{\mathrel{\hbox{\rlap{\hbox{\lower4pt\hbox{$\sim$}}}\hbox{$<$}}}}
\def\gtrsim{\mathrel{\hbox{\rlap{\hbox{\lower4pt\hbox{$\sim$}}}\hbox{$>$}}}}
\def\apj{ApJ}

\def\aap{A\&\hskip-1pt A}
\def\aaps{A\&\hskip-1pt AS}

\title[Bias in Detecting Caustic-Crossing Microlensing Events]
      {On the Intrinsic Bias in Detecting Caustic \\
       Crossings between Galactic Halo and Self-lensing \\
       Events in the Magellanic Clouds}
\author[Cheongho Han]
       {Cheongho Han\\
        Dept.\ of Astronomy \& Space Science, \\
        Chungbuk National University, Chongju, Korea 361-763\\
	cheongho@astronomy.chungbuk.ac.kr}

\date{Accepted
      Received }

\pagerange{\pageref{firstpage}--\pageref{lastpage}}

\begin{document}

\maketitle

\label{firstpage}

\begin{abstract}
In this paper, we investigate the intrinsic bias in detecting caustic 
crossings between Galactic halo and self-lensing events in the Magellanic 
Clouds.  For this, we determine the region for optimal caustic-crossing 
detection in the parameter space of the physical binary separations, 
$\ell$, and the total binary lens mass, $M$, and find that the optimal 
regions for both populations of events are similar to each other.  
In particular, if the Galactic halo is composed of lenses with the 
claimed average mass of $\langle M\rangle\sim 0.5\ M_\odot$, the optimal 
binary separation range of Galactic halo events of 
$3.5\ {\rm AU}\lesssim \ell\lesssim 14\ {\rm AU}$ matches well with 
that of a Magellanic Cloud self-lensing event caused by a binary lens 
with a total mass $M\sim 1\ M_\odot$; well within the mass range of the 
most probable lens population of stars in the Magellanic Clouds.  
Therefore, our computation implies that if the binary fractions and the 
distributions of binary separations of the two populations of lenses are 
not significantly different from each other, there is no strong detection 
bias against Galactic halo caustic-crossing events.
\end{abstract}


\section{Introduction}
From years of monitoring millions of stars located in the Large and 
Small Magellanic Clouds (LMC and SMC), the MACHO (Alcock et al.\ 1997) 
and EROS (Aubourg et al.\ 1993) collaborations have detected $\sim 20$ 
microlensing events.  Among these events, two are caustic-crossing 
binary-lens events (see \S\ 2 for more details about the caustic-crossing 
binary-lens events): one toward the LMC (Bennett et al.\ 1996) and
the other toward the SMC (Alcock et al.\ 1996; Afonso et al.\ 1998; 
Albrow et al.\ 1999; Alcock et al.\ 1998; Udalski et al.\ 1998).
For a caustic-crossing event one can determine the lens proper motion,
$\mu$, from which one can strongly constrain the lens location (Gould 1994; 
Nemiroff \& Wickramasinghe 1994; Witt \& Mao 1994; Peng 1997).  The 
measured lens proper motions for the individual detected binary-lens
events are $\mu\sim 20\ {\rm km\ s}^{-1}$ and $\sim 80\ {\rm km\ s}^{-1}$.
Due to their small values of $\mu$, both lenses are suspected of being 
located within the Magellanic Clouds themselves.  Furthermore, it is 
often hypothesized that stars within the Magellanic Clouds play a dominant 
role as gravitational lenses for events detected toward the Magellanic 
Clouds (Sahu \& Sahu 1998).

The probability of detecting a caustic-crossing binary-lens event 
is strongly dependent on the binary separation in units of the combined 
Einstein ring radius ($b=\ell/r_{\rm E}$: normalized binary separation).
The combined Einstein ring radius is related to the total mass of the 
binary $M$ and its location on the line of sight toward the source star by 
$$
r_{\rm E} = \left( {4GM\over c^2} 
{  D_{ol}D_{ls}\over D_{os}} \right)^{1/2},
\eqno(1)
$$
where $D_{ol}$, $D_{ls}$, and $D_{os}$ are the separations between the 
observer, lens, and source star.  Therefore, the normalized binary 
separation is also related to the total mass and the location of the 
binary lens.  The dominant lens populations (and thus the lens mass) and 
the locations of lenses for Galactic halo and Magellanic Cloud self-lensing 
events are different.  Therefore, even if the binary fractions and the 
distribution of the physical binary separations $f(\ell)$ are similar each 
other, the distributions of the normalized binary separations $f(b)$ of 
the two populations of events might be significantly different from each 
other.  If so, the probability of detecting Galactic halo caustic-crossing 
events will be systematically different from that of self-lensing events 
in the Magellanic Clouds, leading to a detection bias.  Therefore, unless 
it is shown that this type of bias is not important, one cannot conclude 
that MACHOs in the Galactic halo are unlikely to be responsible for the 
events detected toward the Magellanic Clouds.

In this paper, we investigate the intrinsic bias in detecting caustic 
crossings between Galactic halo and self-lensing events in the Magellanic 
Clouds.  For this investigation, we determine the optimal ranges of the 
physical binary separations for caustic crossings.  If the determined
ranges for the two populations of events are similar each other, there will 
be no strong detection bias against Galactic halo caustic-crossing events and 
this therefore will support the hypothesis that an important fraction of 
microlensing events detected toward the Magellanic Clouds are indeed caused 
by lenses in the Magellanic Clouds.

\section{Optimal Binary Separation for Caustic Crossing}

When the lengths are normalized to the combined Einstein ring radius,
the lens equation in complex notation for a binary-lens system is 
represented by 
$$
\zeta = z + {m_{1} \over \bar{z}_{1}-\bar{z}} 
+ {m_{2} \over \bar{z}_{2}-\bar{z}},
\eqno(2)
$$
where $m_1$ and $m_2$ are the mass fractions of individual lenses (and 
thus $m_1+m_2=1$), $z_1$ and $z_2$ are the positions of the lenses, 
$\zeta = \xi +i\eta$ and $z=x+iy$ are the positions of the source and 
images, and $\bar{z}$ denotes the complex conjugate of $z$ (Witt 1990).
The amplification of the binary-lens event is given by the sum of the 
amplifications of individual images, $A_i$, which are given by the Jacobian 
of the transformation (2) evaluated at the image position, i.e.\  
$$
A_i = \left({1\over \vert {\rm det}\ J\vert} \right)_{z=z_i};
\qquad {\rm det}\ J = 1-{\partial\zeta\over\partial\bar{z}}
{\overline{\partial\zeta}\over\partial\bar{z}}.
\eqno(3)
$$
The source positions with infinite amplifications, i.e.\  ${\rm det}\ J=0$, 
form closed curves called caustics.  Whenever a source star crosses a 
caustic, an extra pair of source star images appear (or disappear), 
producing a sharp spike in the light curve (mao \& Paczy\'nski 1991).

\begin{figure}
\epsfysize=10cm
\centerline{\epsfbox{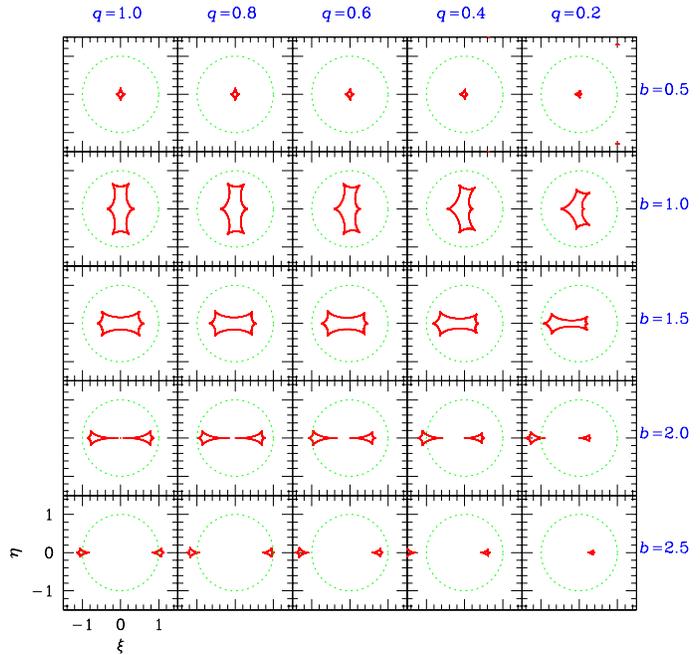}}
\caption{
Caustics (thick solid lines) of gravitational microlensing events 
caused by binary lenses for various normalized binary separations, 
$b$, and mass ratios, $q$.  The positions of lenses are chosen so 
that their center of mass is at the origin.  Both lenses are 
on the $\xi$-axis and the heavier lens is to the right.  The dotted 
circle in each panel represents the combined Einstein ring.
}
\end{figure}

In Figure 1, we present caustics of binary-lens events with 
various values of $b$ and $q$.  In the figure, 
the mass positions $z_1$ and $z_2$ are chosen so that the center 
of mass is at the origin, both lenses are on the $\xi$-axis, 
and the heavier lens is to the right.  From the figure, one finds 
that the caustics take various shapes and sizes depending on the 
values of $b$ and $q$.  As a result, the probability of detecting a 
caustic-crossing binary-lens event is a function of these parameters.

To determine the optimal binary separation for caustic crossing, we 
first compute the caustic-crossing probability as a function of $b$ and 
$q$, $P_{cc}(b,q)$.  To compute $P_{cc}(b,q)$, we must first define a 
binary-lens event.  While a single lens event is defined almost 
unanimously as `a close lens-source encounter within the Einstein ring 
of a lens', there is no clear definition for a binary-lens event. Therefore, 
we define a binary-lens event as `a close lens-source encounter within 
the combined Einstein ring with its center at the center of mass 
of the binary'.  
\footnote{Some binary lenses form their caustics 
outside the combined Einstein ring, e.g.\ a part of the caustics for 
a binary system with $b=2.0$ and $q=0.2$ shown in Figure 1.  According 
to our definition of a binary-lens event, an event with a trajectory 
that passes an outer caustic, but does not enter the combined Einstein 
ring is not a binary-lens event.  However, we note that since outer 
caustics are in general very small compared to the combined Einstein ring, 
the probability $P_{cc}(b,q)$ is not seriously affected by our definition 
of a binary-lens event.}  
With this definition, the caustic-crossing probability is determined 
by computing the ratio of the number of events whose source trajectory 
crosses caustics to the total number of trial trajectories.  The orientation 
angles, $\theta$, of the trial source star trajectory with respect to the 
projected binary axis and the impact parameter (normalized by $r_{\rm E}$), 
$\beta$, are randomly chosen in the ranges $0\leq \theta\leq 2\pi$ and 
$0\leq \beta\leq 1$.  We assume that caustic crossings can be detected as 
long as the source star trajectory crosses any part of the caustics.  The 
upper panel of Figure 2 shows the iso-probability map for caustic-crossing 
in the parameter space of $b$ and $q$.  In the map, contours are drawn 
starting at 10\% and in 10\% intervals.  One finds that caustic crossings 
can happen with an important probability only for some optimal values of 
$b$, and the probability decreases rapidly for binaries with separations 
that are too small or too large. In addition, the probability depends 
weakly on the mass ratio.

\begin{figure}
\epsfysize=10cm
\centerline{\epsfbox{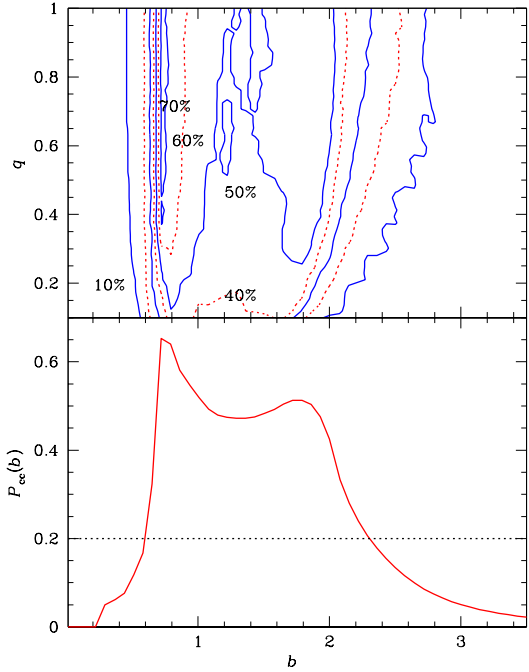}}
\caption{
Upper panel: contours of caustic-crossing probability as a function of 
the normalized binary separation and mass ratio, $P_{cc}(b,q)$.
Contours are drawn at levels starting at 10\% and increasing in steps 
of 10\%.  Lower panel: Caustic crossing probability as a function of only 
the normalized binary separation, $P_{cc}(b)$.  One finds that caustic 
crossings occur with a probability of $P_{cc}(b)\geq 20\%$ when the 
normalized binary separation is in the range $0.6\lesssim b\lesssim 2.3$.
}
\end{figure}

Once $P_{cc}(b,q)$ is computed, the caustic-crossing probability 
as a function of the normalized binary separation is determined by
$$
P_{cc}(b) = \int_0^1 P_{cc}(b,q) f(q) dq,
\eqno(4)
$$
where $f(q)$ is the distribution of binary mass ratios.  To determine 
$P_{cc}(b)$, we assume that $f(q)$ is uniformly distributed. Partially, this
is because $f(q)$ is poorly known, though more importantly because the 
probability $P_{cc}(b,q)$ is weakly dependent on the mass ratio.  In the 
lower panel of Figure 2, we present $P_{cc}(b)$.  One finds that a caustic 
crossing occurs with a probability of $P_{cc}(b)\geq 20\%$ when the 
normalized binary separation is in the range $0.6\lesssim b\lesssim 2.3$.

\section{Bias in Detecting Caustic Crossing Events}
With the optimal range in $b$ we have determined, we then calculate the 
optimal region for caustic crossing in the parameter space of $\ell$ and 
$M$ and illustrate the result in Figure 3.  In the figure, the shaded 
region (enclosed by solid lines) represents the optimal detection region 
for Magellanic Cloud self-lensing events and the unshaded region (enclosed 
by dashed lines) is for Galactic halo events.  The optimal regions are 
determined so that the normalized binary separation for the given values 
of $M$ and $\ell$ lies in the optimal range for $b$, i.e.\ 
$0.6\leq b\leq 2.3$.  The normalized binary separation is related to $M$ 
and $\ell$ by
$$
b (M,\ell,x)
= \left[ {c^2\ell^2\over 4GMD_{os}} 
{1\over x(1-x)}\right]^{1/2};\qquad
x = {D_{ol}\over D_{os}}.
\eqno(5)
$$
Since the Galactic halo's optical depth peaks at $D_{ol}\sim 10\ {\rm kpc}$ 
(Kerins \& Evans 1998), we adopt $x=0.2$ for this population of events.  
For the self-lensing events in the Magellanic Clouds, we adopt the 
average lens-source separation of $\langle D_{ls}\rangle=5\ {\rm kpc}$, 
which is roughly half of the line-of-sight physical depth of the SMC
(Mathewson, Ford, \& Visvanathan 1986; Martin, Maurice, \& Lequex 1989;
Hatzidimitrion et al.\ 1997).

\begin{figure}
\epsfysize=9cm
\centerline{\epsfbox{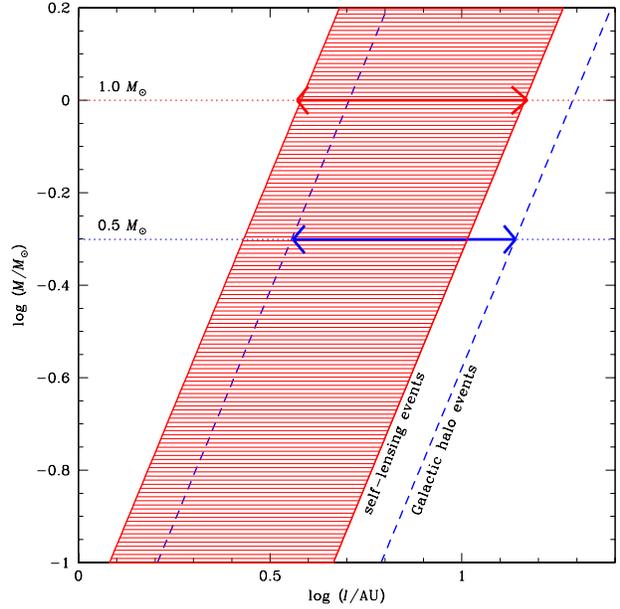}}
\caption{
The region of optimal caustic crossing in the parameter space of the 
physical separations, $\ell$, and total masses of the binaries, $M$.
The shaded region (enclosed by solid lines) represents the optimal region 
for Magellanic Cloud self-lensing events, and the unshaded region (enclosed 
by dashed lines) is for Galactic halo events.  The regions are determined 
so that the normalized binary separation with given values of $\ell$ and 
$M$ lies in the determined optimal range of $0.6\leq b\leq 2.3$.  One 
finds that if the Galactic halo is composed of lenses with an average mass 
of $\langle M\rangle\sim 0.5\ M_\odot$, the optimal range of $\ell$
for the Galactic halo events (represented by the lower arrow) agrees well
with that of a Magellanic Cloud self-lensing event produced by a binary 
lens with a mass of $M\sim 1\ M_\odot$ (represented by the upper arrow).
}
\end{figure}

From the figure, one finds that a large portion of the optimal regions 
for the two populations of events overlap.  For a given lens mass, the 
mean value of the optimal binary separation for Galactic halo events is 
systematically larger than that for Magellanic Cloud self-lensing events.  
However, considering the uncertainties in lens location $x$, this difference 
is not important.  Particularly, if the Galactic halo is composed of 
lenses with the claimed mass of $\langle M\rangle\sim 0.5\ M_\odot$, the 
optimal binary separation range of Galactic halo events is $3.5\ {\rm AU}
\lesssim \ell\lesssim 14\ {\rm AU}$.  This is nearly identical to that of 
the Magellanic Cloud self-lensing events caused by a binary lens with a 
mass $M\sim 1\ M_\odot$, which is well within the mass range of the most 
probable lens population of stars in the Magellanic Clouds.  Therefore, 
our computation implies that there is no strong detection bias against
Galactic halo caustic-crossing events and supports the hypothesis that 
a significant fraction of events detected toward the Magellanic Clouds 
are caused by lenses in the Magellanic Clouds themselves.

\section*{Acknowledgments}
We would like to thank P.\ Martini for carefully reading the manuscript.

\end{document}